# A Requirements Modeling Language for the Component Behavior of Cyber Physical Robotics Systems

Jan Oliver Ringert, Bernhard Rumpe, and Andreas Wortmann

RWTH Aachen University, Software Engineering, Aachen, Germany
{ringert,rumpe,wortmann}@se-rwth.de

**Abstract.** Software development for robotics applications is a sophisticated endeavor as robots are inherently complex. Explicit modeling of the architecture and behavior of robotics application yields many advantages to cope with this complexity by identifying and separating logically and physically independent components and by hierarchically structuring the system under development. On top of component and connector models we propose modeling the requirements on the behavior of robotics software components using I/O$^\omega$ automata [29]. This approach facilitates early simulation of requirements model, allows to subject these to formal analysis and to generate the software from them. In this paper, we introduce an extension of the architecture description language MontiArc to model the requirements on components with I/O$^\omega$ automata, which are defined in the spirit of Martin Glinz' Statecharts for requirements modeling [10]. We furthermore present a case study based on a robotics application generated for the Lego NXT robotic platform.

> *"In der Robotik dachte man vor 30 Jahren, dass man heute alles perfekt beherrschen würde", Martin Glinz [38]*

## 1 Introduction

Robotics is a field of Cyber-Physical Systems (CPS) which yields complex challenges due to the variety of robots, their forms of use and the overwhelming complexity of the possible environments they have to operate in. Software development for robotics applications is still at least as complex as it was 30 years ago: even a simple robot requires the integration of multiple distributed software components. Successful robotics applications – for example the RoboCup[1] contributions to robotic soccer and service robotics – are usually the joint effort of teams of domain experts. The results are integrated into experimental, monolithic and hardly adaptable or reusable platforms [22,33].

While software reuse in robotics has been intensively pursued within the last five years [6,5], most approaches focus on some form of component-based software engineering (CBSE) reusing binary components [26,24], whereas modeling techniques have been proven useful in several other domains [27,1] to reduce the "accidental complexities" [9] arising from the gap between problem domain and implementation domain.

---
[1] http://www.robocup.org/

Modeling the architecture of robots yields many advantages to cope with these complexities. Architecture descriptions allow to identify and separate logically and physically independent components and to hierarchically structure the system under development. Models are language-agnostic, can be provided by domain experts, are highly reusable and may be better suited for formal analysis than code written in general purpose languages (GPLs). Furthermore, executable models may be transformed into the target system using software engineering knowledge embodied in generation tools, liberating the domain experts from becoming software engineering experts. We use the architecture description language (ADL) MontiArc [13] to model and simulate distributed, hierarchically decomposed, service robotics systems. This means we can model robots as Component and Connector (C&C) architectures, where a component is a unit performing computations and the information flow between components is defined by unidirectional connectors between typed ports of the components. MontiArc enforces stable component interfaces, thus the components and their requirements may be developed independently, minimizing integration effort.

Statechart descriptions are an amenable mechanism to model interactive systems [14,10]. We use I/O$^\omega$ automata to model requirements for the runtime behavior of components and subsystems. Following the idea to describe requirements in Statechart languages "as simple as possible, as rich as needed" [10], we present a variant of I/O$^\omega$ automata [29] to model requirements on the input and output reaction of components in the architecture of robotic systems. This approach to requirements modeling yields several advantages:

- I/O$^\omega$ automata allow underspecification in two forms: (1) incompleteness of triggers to only regulate the reaction to inputs of interest and (2) nondeterministically overlapping triggers to restrict possible behavior as desired.
- Incomplete I/O$^\omega$ automata can be composed as well as refined to more detailed and complete behavior.
- The use of a Component & Connector architecture description language makes communication and dependencies explicit in the models.
- Due to the embedding into MontiArc components, requirements may be modeled and simulated independently, incrementally and bottom-up by different domain experts, thus facilitating the evolutionary development of the requirements models as proposed in [34,11].
- The models may further be validated automatically by means of model transformation into other formalisms. A translation of I/O$^\omega$ automata to the model checker Mona [8] for analysis tasks, e.g., refinement checking is presented in [16].

We illustrate the requirements modeling using an example coffee service application in Section 2, before we describe the semantics of the I/O$^\omega$ automata modeling language in Section 3. In Section 4, we show how models of this language may be utilized to generate executable implementations, discuss related work in Section 5 and finally conclude our approach to requirements modeling for robotic systems in Section 6.

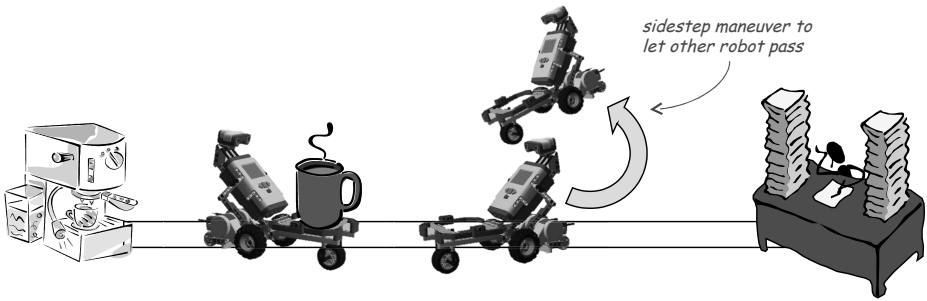

**Fig. 1.** A scenario where one coffee robot encounters another coffee robot in opposite direction on a common path.

## 2 The Coffee Service Example

We present a simple example of a coffee service robot. The objective of the robot is to drive to a coffee machine and pick up coffee that it delivers to coffee drinkers at different locations. To deliver coffee the robot follows marked lines on the floor for orientation. If a coffee fetching robot encounters another one on the same line but in opposite direction it either has to wait for the second robot to give way or has to give way itself (sidestep maneuver). This situation and the sidestep maneuver are shown in Fig. 1.

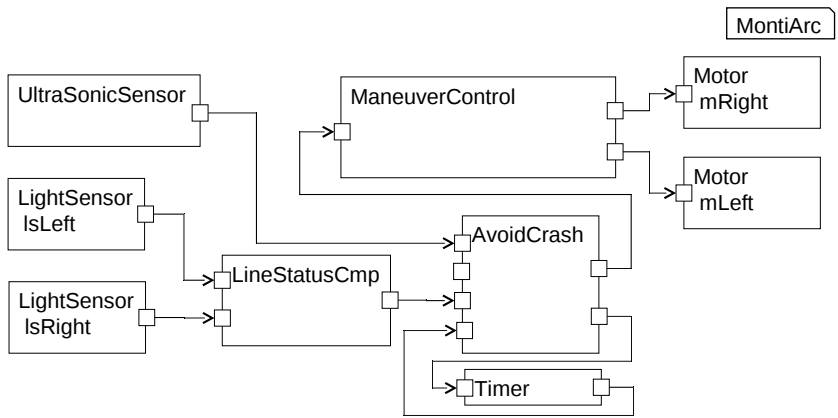

**Fig. 2.** A partial overview of the architecture of the coffee fetching robot presented as a MontiArc model. Sensors (ultrasonic and two light sensors) are aligned to the left and actuators (two motors) are aligned to the right.

The engineering team already created the partial C&C architecture of the coffee fetching robot shown in Fig. 2. The main component of the architecture is the

`AvoidCrash` component. It receives input from an ultrasonic sensor and information from component `LineStatusComp` whether the two light sensors have correctly captured the line to follow. Other robots are detected via the ultrasonic sensor by measuring the distance to the next object.

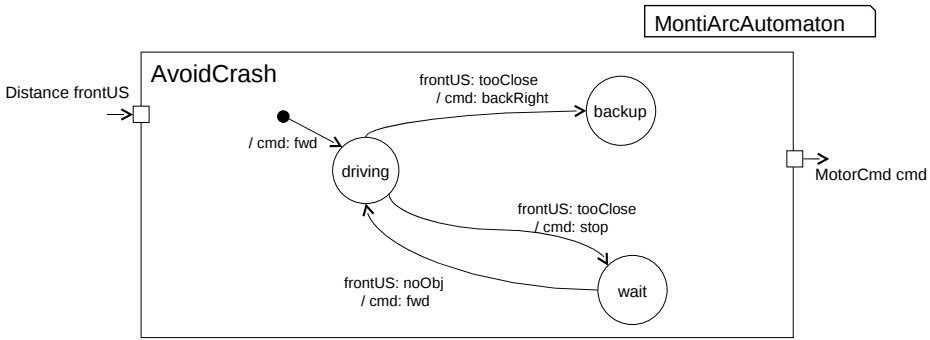

**Fig. 3.** Initial and incomplete model of the input and output behavior of component `AvoidCrash` given as an I/O$^\omega$ automaton.

An engineer has described requirements for the input and output behavior of component `AvoidCrash` as the I/O$^\omega$ automaton shown in Fig. 3. Component `AvoidCrash` has to start in an initial state `driving` and to output the message `fwd` on its outgoing port `cmd` of type `MotorCmd`. This triggers the engines to drive forward. The possible messages that can be sent on the port `cmd` are the values of the enumeration type `MotorCmd` shown in the class diagram in Fig. 4. If the component receives a `tooClose` message in state `driving` it should either send the motor command `backRight` to back up away from the line in a right curve to let the second robot pass or it should send the motor command `stop` and wait until the ultra sonic sensor reports that the robot detected earlier has disappeared.

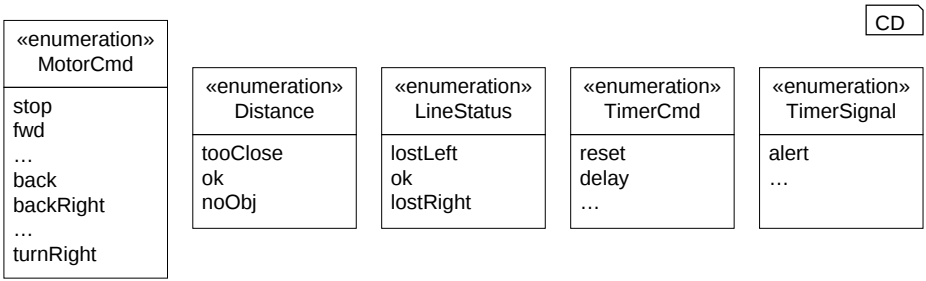

**Fig. 4.** UML/P class diagram of the enumerations used as message types on input and output ports of the `AvoidCrash` component.

The requirement specification in Fig. 3 is rather incomplete. Over time, the engineernig team needs to detail the required behavior of the robot during the sidestep maneuver. As more knowledge of the system under development is aquired, the early requirements are typically refined into more detailed behavior descriptions. Ideally, continuous refinement of requirements leads to an implementation.

A refined version of the I/O behavior definition of component `AvoidCrash` is shown in Fig. 5. This version details the handling of the sidestep maneuver: the component sets a timer by sending the `delay` command to a timer component. The timer is expected to respond with an `alert` message once the delay has elapsed. The robot then has to stop motors at its remote position and wait until the second robot has cleared the line. It then should issue a `fwd` command to the motors to return to the line. This description of the system's input and output is detailed enough to simulate exemplary system runs.

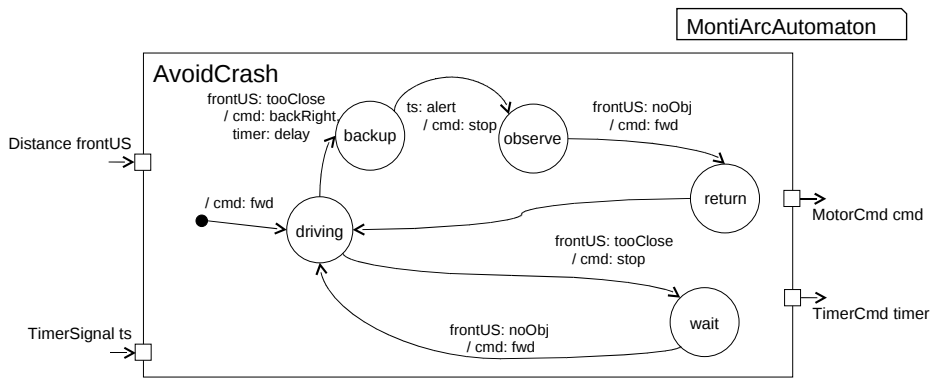

**Fig. 5.** An I/O specification for component `AvoidCrash` with additional details about the sidestep maneuver once an obstacle is detected.

The examples demonstrate some features of the message passing semantics of I/O$^\omega$ automata for describing required I/O behavior. This model allows communication between components on an event or command level. An example for this high-level communication is sending the `fwd` command only once and not in every execution cycle in that the robot should be driving forward. I/O$^\omega$ automata however also allow continuous sending of data messages as for example done by the ultrasonic sensor on port `frontUS`. These values are only read on transitions when necessary.

Working towards a behavior implementation based on I/O$^\omega$ automata, the coffee fetcher development team decided for a resolution of the nondeterminism between the sidestep and the stop and wait maneuver. If the robot carries a fresh cup of coffee the second robot has to sidestep (saving time for the robot that delivers the coffee). Please note that the target locations are arranged in a tree like structure around the coffee machine and thus no two robots with fresh coffee will ever face each other in opposite directions. The refined version of the `AvoidCrash` component is shown in Fig. 6.

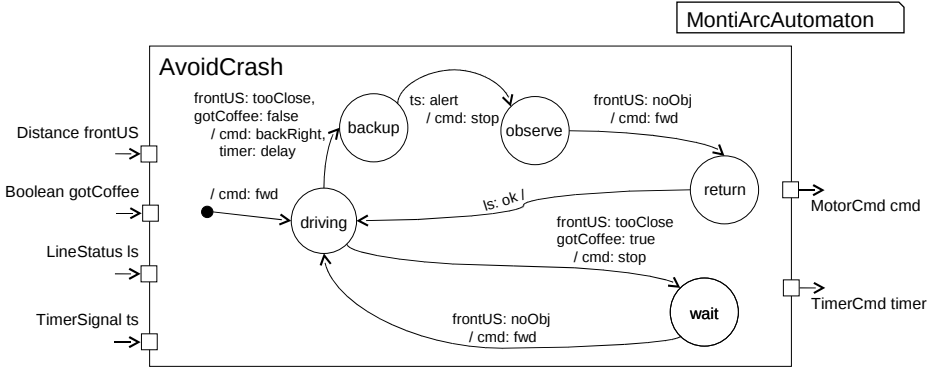

**Fig. 6.** Updated version of component `AvoidCrash` with nondeterminism between sidestep and stop and wait action resolved.

Furthermore the team decided that when driving back from the remote position the message that the line was found (value `ok` on port `ls` of type `LineStatus`) should trigger the transition to the initial driving state.

The examples from Fig. 5 and Fig. 6 show that messages on multiple input and output ports can be received or sent from single transitions. This feature can, e.g., be used if transitions are guarded by values read from sensors or dependent on an external state like the value on port `gotCoffee`.

The first version of the `AvoidCrash` component from Fig. 3 is modeled with an incomplete interface and only a few required output messages in response to its local state and the limited inputs known. The last version shown in Fig. 6 is much more detailed and many decisions are fixed. All versions of I/O$^\omega$ automata can easily be used for code generation and simulation. However, if the automaton in partially underspecified, choices need to be made [29]. The more detailed, the more functionality we ca test, simulate or use. We consider it a benefit of the language that an engineer can easily sketch requirements as partial models but also create detailed behavior models.

## 3   I/O$^\omega$ Automata Modeling Language

We use the architecture description language (ADL) MontiArc [13] to model robotic systems. The top level elements offered by MontiArc to describe interactive distributed systems are components and connectors following the definitions of Medvidovic and Taylor [21,37]: components encapsulate subsets of the systems functionality and regulate access via explicitly defined interfaces. Connectors allow and regulate the interaction of components. Component interfaces in MontiArc are sets of directed (either input or output) typed ports. The type associated with a port determines the possible messages a component may receive or send on that port.

In MontiArc and its underlying semantic model FOCUS [4] a component is either atomic and its behavior is defined explicitly or a component is decomposed and its behavior is defined solely based on the structural composition of the behaviors of its sub-components. Models of decomposed components represent configurations of subsystems (relations between components and connectors). MontiArc cleary distinguishes between the definition of a component and its instantiation, but allows to efficiently define and instantiate components where desired. It also supports powerful typing, instantiation, and parametrization mechanisms as described in [13] that allow, e.g., the definition of generic components and reuse by instantiation and parametrization.

The ADL MontiArc does itself not provide a language to model the input and output behavior of its components explicitly (implicit definitions are possible using embedded constraint languages). We have thus extended MontiArc by embedding I/O$^\omega$ automata as form of component definitions. Our implementation of this DSL using the MontiCore framework is called MontiArcAutomaton. We use the terms I/O$^\omega$ automaton and MontiArcAutomaton automaton interchangeably in this paper.

### 3.1 MontiArcAutomaton Language Elements

The purpose of the modeling language MontiArcAutomaton is to model I/O$^\omega$ automata that describe requirements on the input and output behavior of software components. State-based descriptions are an amenable mechanism to model requirements for interactive systems [14,10].

The language elements added to MontiArc components in order to model component behavior are local variables, states, and transitions. Variables can be used by automata to store intermediate values to, e.g., implement counters or timers. A transition connects a source and a target state and has an optional guard, input block, and output block. Variables, states and transitions are only visible inside a component and all communication between components is made explicit via ports and connectors. Via local variables to counter state space explosion, via the rich composition operators of ADLs, and via explicit communication mechanisms we believe to make our requirements language accessible, easy to understand, and yet powerful enough as motivated in [10].

The input block of a transition defines patterns of events and messages received on incoming ports or stored in the local variables of the component that together with the guard activate the transition. A guard is a predicate over the messages on input ports and values stored in local variables. Guards in MontiArcAutomaton can be specified using OCL/P [30,31] our MontiCore implementation of OCL.

A component's reaction to an input is specified in the output block of a transition. The output block is a set of pairs of output ports and the messages or streams of messages that are sent as a reaction to the input. The output block also may contain assignments to update the local variables of a component.

In the language design of MontiArcAutomaton we deliberately omitted an action language known, e.g., from UML [12]. We believe that requirements for the behavior of interactive components are best described and understood on an abstract and not too implementation oriented event- and message-based level while an action language often degrades into a programming language itself.

Please note that when implementing a Cyber-Physical System each component should have exactly one corresponding implementation. In our case we can have multiple I/O$^\omega$ automata denoting requirements on the behavior of a single atomic component.

We furthermore can have I/O$^\omega$ automata describing the behavior of a decomposed component and thus imposing requirements on the implementation of its sub components.

### 3.2  Semantics of I/O$^\omega$ Automata

The semantics of I/O$^\omega$ automata is described as sets of stream processing functions [3,4,28]. Each stream processing function (SPF) describes one realizable behavior (i.e., an implementation) of the desired system. The SPF corresponding to a MontiArc component maps one input stream bundle to one output stream bundle. The input stream bundle contains streams for each input port of the component and the output stream bundle contains streams for each output port of the component. For a formal definition of I/O$^\omega$ automata semantics see [29] and the MontiArcAutomaton website [17].

When defining requirements for the behavior of an interactive system we typically allow many different concrete implementations. To be able to define these requirements in way that is not too restrictive and too implementation oriented I/O$^\omega$ automata offer various features to model underspecified behavior and express partial knowledge by allowing nondeterministic choice between enabled transition as well as completely unspecified behavior when no transition is enabled.

### 3.3  Language Features for Partial Knowledge and Requirements

Partiality of models and ways to express underspecified behavior are important when modeling requirement scenarios [35]. Modeling requirements and partial knowledge available is supported by I/O$^\omega$ automata in multiple ways. These mechanisms can be used to express the level of confidence of the completeness of the behavior and behavior requirements modeled. The most complete level is achieved when the I/O$^\omega$ automaton handles all input messages of interest and defines according outputs in all states. The I/O$^\omega$ automata language allows to introduce nondeterminism by modeling multiple transitions activated for the same input to describe alternative behaviors.

The set of states and transitions of an automaton can be interpreted as incomplete. This means that the transitions shown only denote required behavior. Nothing is specified about the I/O behavior of the system if no transition is activated. An implementation or refinement of the required behavior can add states and transitions where previously no behavior was defined.

Another notion of incompleteness is when the set of states is completely known but further transitions might exist. This requires again any implementation to behave according to the defined input and output behavior. In case an undefined input (combination) is read, the output of the component is arbitrary but the component has to continue computation from its current state. Thus state changes can only happen via transitions defined in the automaton by the requirements engineer.

Note that state and transition complete I/O$^\omega$ automata are not required to explicitly handle inputs from all incoming ports. In our semantics described on the MontiArcAutomaton website [17] this simply means that all possible inputs on the ports not considered would have no effect on the modeled behavior and are thus ignored.

## 4  Code Generation and Case Study

We use models of the MontiArcAutomaton [17] language to describe the required runtime behavior of systems modeled as component architectures. Based on the completions described in Section 3.3 we can utilize MontiCore's template based code generation toolchain [32] to generate executable component architectures directly from the requirements specification. MontiCore's compositional approach to language development allows to generate code for different target platforms easily [18]. We are applying MontiCore code generators for I/O$^\omega$ automata to first model the requirements of the coffee service illustrated in Section 2 and subsequently generate executable code from detailed behavior descriptions in our current service robotics course.

The participants use Lego NXT kits and the Lego Java Operating System[2] (leJOS) to build three robots (cup dispenser, coffee fetcher, coffee preparing robot) working together to provide the coffee service illustrated in Section 2 over the course of a semester. We provided the MontiArcAutomaton language, several MontiArc components wrapping the sensors and actuators provided by leJOS, initial requirements as user stories and a source code generator targeting leJOS. The students work in groups on different parts of the system using I/O$^\omega$ automata to model the requirements of the components they develop. In doing so, they evolve sequences of requirements models that, bottom-up, converge against the desired system behavior.

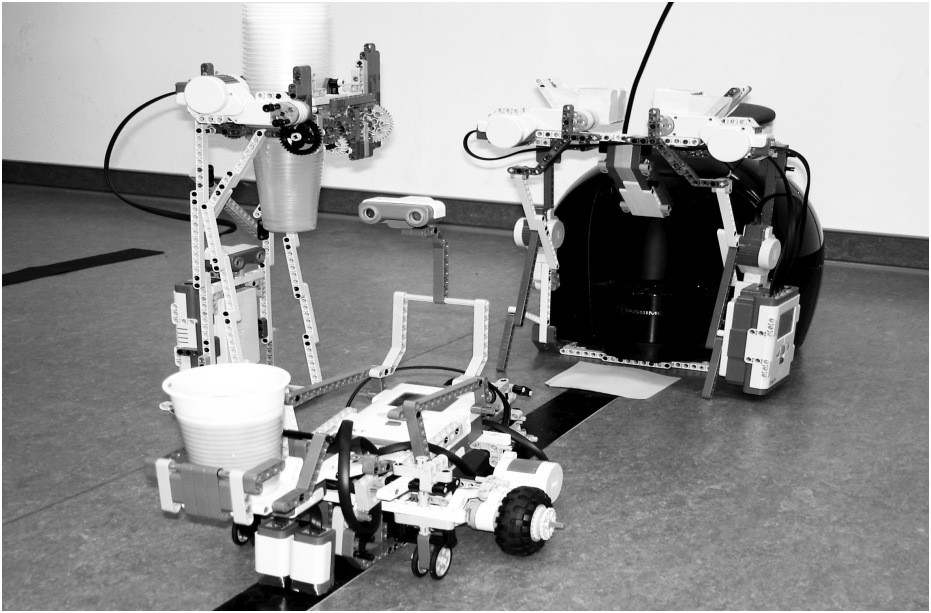

**Fig. 7.** The three Lego NXT robots providing the coffee service.

---

[2] http://lejos.sourceforge.net/

The source code generation uses the framework introduced in [32], i.e., we have provided templates and template operators capturing the transformation of MontiArcAutomaton models to leJOS compatible Java code. Therefore, we imposed a three layer architecture (sensors, logics, actuators) on the participants, depending on the leJOS API only in the sensors and actuators layers. This easily allows to subject the business logics layer to simulation and to port it to other robots. The target implementation details (e.g., programming languages, robot APIs, libraries) only depend on the generator templates, thus it is straightforward to generate for different targets (developing the prototype of a code generator for leJOS Java took us two days).

## 5   Related Work

Research in robotics software reuse produced to a number of CBSE frameworks [7,2,5,20,24,33] and other approaches to modular robot architectures [39,26,19], while modeling robot behavior has received less attention [23,40,25]. Integrated architecture and behavior modeling is even less common. The approach proposed in [33] to model components with embedded partial statecharts to be filled by the developer follows a "freedom from choice" approach: the developer, for example, is forced to use statecharts for behavior modeling and C++ as target language. MontiArc instead allows to embed arbitrary domain specific languages developed with MontiCore into the components (e.g., Java/P [32]) and to generate source code for arbitrary target platforms. Besides the research in dedicated robotics modeling, there are several other modeling languages and frameworks available to model both system architecture and behavior requirements.

The System Modeling Language[3] (SysML) is a variant of UML which features, among others, the internal block diagrams ADL and state machine diagrams. While SysML can be used to model overall system requirements a priori [15,36] there yet are not approaches to incrementally model the behavior requirements of CPS in a form, such that the models may be subjected to simulation at each instant. While it further is possible to use statecharts instead of I/O$^\omega$ automata, this would require to modify the semantics of until these result in a more complex form of I/O$^\omega$ automata – which we, following [10], want to avoid.

## 6   Conclusion

We have shown how the embedding on I/O$^\omega$ automata into MontiArc components can be used to evolutionary model partial requirements on the behavior of a certain type of CPS. The resulting MontiArcAutomaton language allows to model both architecture and behavior of the system under development incrementally, bottom-up and in a distributed manner. We have illustrated syntax and semantics of the MontiArcAutomaton language and discussed that these can be used to easily develop code generators for different platforms.

---

[3] http://www.sysml.org/

Interesting directions for future work include an evaluation of the usefulness of the modeling language for describing the requirements and developing the control logic of robotic systems (partially done based on the current student lab described in Section 4). We are also interested in extending our initial analysis framework for combining multiple models and analyzing the evolution of requirements as well as to automatically discover contradicting (sets of) scenarios.

**Acknowledgments**

We thank Arne Haber for fruitful language discussions and MontiArc support and Dennis Kirch for discussions on language features and MontiArcAutomaton context conditions.

J.O. Ringert is supported by the DFG GK/1298 AlgoSyn.

## References

1. Berger, C., Rumpe, B.: Autonomous Driving - 5 Years after the Urban Challenge: The Anticipatory Vehicle as a Cyber-Physical System. In: Goltz, U., Magnor, M., Appelrath, H.J., Matthies, H.K., Balke, W.T., Wolf, L. (eds.) Proceedings of the INFORMATIK 2012 (2012)
2. Brooks, A., Kaupp, T., Makarenko, A., Williams, S., Oreback, A.: Towards component-based robotics. In: Intelligent Robots and Systems, 2005.(IROS 2005). 2005 IEEE/RSJ International Conference on. pp. 163–168. IEEE (2005), http://ieeexplore.ieee.org/xpls/abs\_all.jsp?arnumber=1545523
3. Broy, M., Dederich, F., Dendorfer, C., Fuchs, M., Gritzner, T., Weber, R.: The Design of Distributed Systems - An Introduction to FOCUS. Tech. rep., TUM-I9202, SFB-Bericht Nr. 342/2-2/92 A (1993)
4. Broy, M., Stølen, K.: Specification and Development of Interactive Systems. Focus on Streams, Interfaces and Refinement. Springer Verlag Heidelberg (2001)
5. Brugali, D., Brooks, A., Cowley, A., Côté, C., Domínguez-Brito, A., Létourneau, D., Michaud, F., Schlegel, C.: Trends in Component-Based Robotics. In: Brugali, D. (ed.) Software Engineering for Experimental Robotics, Springer Tracts in Advanced Robotics, vol. 30, chap. 8, pp. 135–142. Springer, Berlin, Heidelberg (2007), http://dx.doi.org/10.1007/978-3-540-68951-5\_8
6. Brugali, D., Salvaneschi, P.: Stable Aspects In Robot Software Development. International Journal of Advanced Robotic Systems 3 (2006), http://www.intechopen.com/articles/show/title/stable\_aspects\_in\_robot\_software\_development
7. Bruyninckx, H.: Open robot control software: the OROCOS project. Proceedings 2001 ICRA. IEEE International Conference on Robotics and Automation (Cat. No.01CH37164) 3, 2523–2528 (2001), http://ieeexplore.ieee.org/lpdocs/epic03/wrapper.htm?arnumber=933002
8. Elgaard, J., Klarlund, N., Møller, A.: MONA 1.x: new techniques for WS1S and WS2S. In: Computer-Aided Verification, (CAV '98). LNCS, vol. 1427, pp. 516–520. Springer-Verlag (1998)
9. France, R., Rumpe, B.: Model-driven development of complex software: A research roadmap. In: Future of Software Engineering 2007 at ICSE. pp. 37–54 (2007)

10. Glinz, M.: Statecharts for requirements specification - as simple as possible, as rich as needed. In: International Conference on Software Engineering (ICSE) 2002. ACM Press (2002)
11. Glinz, M., Seybold, C., Meier, S.: Simulation-driven creation, validation and evolution of behavioral requirements models. In: Conrad, M., Giese, H., Rumpe, B., Schätz, B. (eds.) MBEES. Informatik-Bericht, vol. 2007-1, pp. 103–112. TU Braunschweig, Institut für Software Systems Engineering (2007)
12. Object Management Group: Unified Modeling Language: Superstructure Version 2.4.1. http://www.omg.org/spec/UML/2.4.1/ (2012), accessed 10/12
13. Haber, A., Ringert, J.O., Rumpe, B.: Montiarc - architectural modeling of interactive distributed and cyber-physical systems. Tech. Rep. AIB-2012-03, RWTH Aachen (february 2012), http://aib.informatik.rwth-aachen.de/2012/2012-03.pdf
14. Harel, D.: Statecharts: A visual formalism for complex systems. Sci. Comput. Program. 8(3), 231–274 (1987)
15. Huang, E., Ramamurthy, R., McGinnis, L.F.: System and simulation modeling using sysml. In: Proceedings of the 39th conference on Winter simulation: 40 years! The best is yet to come. pp. 796–803. WSC '07, IEEE Press, Piscataway, NJ, USA (2007), http://dl.acm.org/citation.cfm?id=1351542.1351687
16. Kirch, D.: Analysis of Behavioral Specifications for Distributed Interactive Systems with MONA. Bachelor Thesis, RWTH Aachen University (2011)
17. Kirch, D., Ringert, J.O., Rumpe, B., Wortmann, A.: MontiArcAutomaton Project Website. http://www.se-rwth.de/materials/ioomega, accessed 10/2012
18. Krahn, H., Rumpe, B., Völkel, S.: MontiCore: a framework for compositional development of domain specific languages. STTT 12(5), 353–372 (2010)
19. Lütkebohle, I., Wachsmuth, S.: Requirements and a Case-Study for SLE from Robotics: Event-oriented Incremental Component Construction. Workshop on Software-Language-Engineering for Cyber-Physical Systems (2011), http://www.user.tu-berlin.de/komm/CD/paper/061423.pdf
20. Makarenko, A., Brooks, A., Kaupp, T.: Orca : Components for Robotics. In: Brugali, D. (ed.) Software Engineering for Experimental Robotics, pp. 231–251. Springer Berlin/Heidelberg (2007)
21. Medvidovic, N., Taylor, R.: A Classification and Comparison Framework for Software Architecture Description Languages. IEEE Transactions on Software Engineering (2000)
22. Mosterman, P.J.: Elements of a Robotics Research Roadmap: A Model-Based Design Perspective (2009)
23. Murray, J.: Specifying agents with uml in robotic soccer. In: Proceedings of the first international joint conference on Autonomous agents and multiagent systems: part 1. pp. 51–52. AAMAS '02, ACM, New York, NY, USA (2002), http://doi.acm.org/10.1145/544741.544756
24. Niemueller, T., Ferrein, A., Beck, D., Lakemeyer, G.: Design Principles of the Component-Based Robot Software Framework Fawkes, Lecture Notes in Computer Science, vol. 6472, chap. NFB+10, pp. 300–311. Springer, Darmstadt, Germany (2010)
25. Niemueller, T., Ferrein, A., Lakemeyer, G.: A Lua-based Behavior Engine for Controlling the Humanoid Robot Nao. In: Proc. of RoboCup Symposium 2009. Graz, Austria (2009)
26. Quigley, M., Gerkey, B., Conley, K., Faust, J., Foote, T., Leibs, J., Berger, E., Wheeler, R., Ng, A.: ROS: an open-source Robot Operating System. In: ICRA Workshop on Open Source Software. No. Figure 1 (2009), http://pub1.willowgarage.com/~konolige/cs225B/docs/quigley-icra2009-ros.pdf
27. Raistrick, C.: Applying mda and uml in the development of a healthcare system. In: Jardim Nunes, N., Selic, B., Rodrigues da Silva, A., Toval Alvarez, A. (eds.) UML Modeling

Languages and Applications, Lecture Notes in Computer Science, vol. 3297, pp. 203–218. Springer Berlin / Heidelberg (2005)
28. Ringert, J.O., Rumpe, B.: A Little Synopsis on Streams, Stream Processing Functions, and State-Based Stream Processing. International Journal of Software and Informatics 5(1-2), 29–53 (July 2011)
29. Rumpe, B.: Formale Methodik des Entwurfs verteilter objektorientierter Systeme. Doktorarbeit, Technische Universität München (1996)
30. Rumpe, B.: Modellierung mit UML. Xpert.press, Springer Berlin, 2nd edn. (September 2011)
31. Schindler, M.: Eine Werkzeuginfrastruktur zur Agilen Entwicklung mit der UML/P. Ph.D. thesis, RWTH Aachen (2011), to appear
32. Schindler, M.: Eine Werkzeuginfrastruktur zur agilen Entwicklung mit der UML/P. Aachener Informatik-Berichte, Software Engineering, Band 11, Shaker Verlag (2012)
33. Schlegel, C., Steck, A., Lotz, A.: Model-Driven Software Development in Robotics : Communication Patterns as Key for a Robotics Component Model. In: Chugo, D., Yokota, S. (eds.) Introduction to Modern Robotics. iConcept Press (2011), http://www.iconceptpress.com/01/site/publication.downloadPaper.php?paperID=101215045543
34. Seybold, C., Meier, S., Glinz, M.: Evolution of requirements models by simulation. In: IWPSE. pp. 43–48. IEEE Computer Society (2004)
35. Seybold, C., Meier, S., Glinz, M.: Scenario-driven modeling and validation of requirements models. In: Whittle, J., Geiger, L., Meisinger, M. (eds.) SCESM. pp. 83–89. ACM (2006)
36. Soares, M.S., Vrancken, J.L.M.: Requirements specification and modeling through sysml. pp. 1735–1740 (2007)
37. Taylor, R.N., Medvidovic, N., Dashofy, E.M.: Software Architecture: Foundations, Theory, and Practice. John Wiley and Sons, Inc., 1 edn. (2009)
38. Universität Zürich: Universität Zürich Jahresbericht 2010. http://www.uzh.ch/about/portrait/annualreport/Jahresbericht_2010.pdf (2010), accessed 10/2012
39. Volpe, R., Nesnas, I., Estlin, T., Mutz, D., Petras, R., Das, H.: The CLARAty architecture for robotic autonomy. In: Aerospace Conference, 2001, IEEE Proceedings. vol. 1, pp. 1/121–1/132 vol.1 (2001)
40. Williams, B.C., Ingham, M.D., Chung, S.H., Elliott, P.H.: Model-based programming of intelligent embedded systems and robotic space explorers. In: Proceedings of the IEEE: Special Issue on Modeling and Design of Embedded Software. pp. 212–237 (2003)